\documentclass[prb,twocolumn,showpacs,preprintnumbers,amsmath,amssymb]{revtex4}
\usepackage{amsthm,amscd,graphicx,color,colordvi,bm,nicefrac,psfrag}

\newcommand{\vect}[1]{\mathbf{#1}}
\newcommand{\hvect}[1]{\hat{\mathbf{#1}}}
\newcommand{\vectsym}[1]{\bm{#1}}
\renewcommand{\tensor}[1]{\bm{#1}}
\newcommand{\pdif}[2]{\frac{\partial #1}{\partial #2}}
\renewcommand{\exp}[1]{\text{e}^{#1}}
\newcommand{\Fourier}[1]{\widetilde{#1}}
\newcommand{\BesselJ}{\mathcal{J}}
\newcommand{\betaP}{\beta^{\text{P}}}

\newcommand{\rhoF}{\Fourier{\rho}}
\newtheorem{theorem}{Theorem}

\begin{document}

\preprint{cond-mat/0610492}

\title{%
Stress-free states of continuum dislocation fields:
Rotations, grain boundaries, and the Nye dislocation density tensor}

\author{Surachate Limkumnerd}
\email{s.limkumnerd@rug.nl}
\affiliation{%
Zernike Institute for Advanced Materials, Nijenborgh 4,
University of Groningen, Groningen 9747AG, The Netherlands
}

\author{James P. Sethna}
\homepage{http://www.lassp.cornell.edu/sethna/sethna.html}
\affiliation{%
Laboratory of Atomic and Solid State Physics, Clark Hall,\\
Cornell University, Ithaca, NY 14853-2501, USA
}

\date{\today}

\begin{abstract}
We derive general relations between grain boundaries, rotational
deformations, and stress-free states for the mesoscale continuum Nye
dislocation density tensor. Dislocations generally are associated with
long-range stress fields. We provide the general form for dislocation
density fields whose stress fields vanish. We explain that a
grain boundary (a dislocation wall satisfying Frank's formula) has
vanishing stress in the continuum limit. We show that the
general stress-free state can be written explicitly as a (perhaps
continuous) superposition of flat Frank walls. We show that the stress-free
states are also naturally interpreted as configurations generated by a
general spatially-dependent rotational deformation. Finally, we propose a
least-squares definition for the spatially-dependent rotation field of a
general (stressful) dislocation density field. 

\end{abstract}


\maketitle

\section{Introduction}
\label{sec:Introduction}
The crystalline phase breaks two continuous symmetries: translational
invariance and rotational invariance. Dislocations are the topological
defect associated with broken translational symmetry. Rotational distortions
in crystals relax into wall structures formed from arrays of dislocations.%
  \footnote{Disclinations, the topological defect suggested by homotopy
  theory applied to the broken rotational symmetry, are 
  forbidden in bulk crystals because the broken translational symmetry
  makes the long-range rotational distortions too costly in energy;
  disclinations are screened by dislocations arranged into grain boundaries.}
It would seem natural to describe the formation and evolution of these
mesoscale dislocation structures using a continuum dislocation density
theory. Individual dislocations are associated with long-range stress
fields, and the dislocation evolution and structure formation is
strongly constrained by the need to screen these stresses.
In the 1950s, a number of authors~\cite{Eshe56,Kron58,Kron81,%
Kose62,Mura63,Mura91} (building on ideas from differential geometry) 
developed such a continuum description involving the coarse-grained
net topological charge of the dislocations in a region, organized 
into dislocation density tensor named for Nye. Since the long-range
stress fields and the rotational distortions only depend upon the net
dislocation density, it is natural for us to use this order parameter to
describe the connections between rotations, dislocations, and stress.
In this manuscript, we provide a systematic mathematical analysis
of the relations between rotational deformations, domain walls, and dislocation
stress within the framework of Nye's continuum dislocation density tensor.

In equilibrium, a large crystal with boundary conditions imposing a rotational
deformation will relax by forming a few grain boundaries---sharp walls
formed by dislocation arrays, separating perfect crystalline
grains, with elastic stress confined locally near the grain boundaries.%
  \footnote{For example, a cube of side $L$ bent by a small misorientation
  angle $\theta$ has an elastic energy $\propto L^3 \theta^2$ which 
  can be relieved by introducing a grain boundary of energy 
  $\propto L^2 b\, \theta \log(1/\theta)$ where $b$ is the length of the
  Burgers vector (a lattice constant). The grain boundary hence forms when the
  bending angle is larger than $\theta_c \sim (b/L) \log(L/b)$, 
  which vanishes as $L$ gets large. Interestingly, the net amplitude
  of deformation at the midpoint of the bent cube is 
  $(L/2) \theta_c \sim b \log(L/b) \sim b$ at the point where the grain
  boundary forms---just a few lattice constants. Hence elastic 
  deformations at the boundary larger than a few lattice constants will
  form grain boundaries~\cite{SethHuan92}.}
In the continuum theory, the region of local elastic stress vanishes,
and grain boundaries are described as $\delta$-function dislocation densities
with zero stress.
We analyze stress-free dislocation walls in the continuum theory in 
section~\ref{sec:FlatWallDecomposition}, and connect them with Frank's 
formula~\cite{Fran50} for the dislocation content of grain boundaries in
appendix~\ref{app:FrankCondition}.

At high temperatures, an initially disordered or microscopically deformed
material will approach equilibrium through the formation and coarsening
of polycrystals---the rotation gradients are confined to sharp grain boundaries,
separating local regions of different orientations.
Again, the strain fields in a polycrystal are confined to small regions
around the grain boundaries. In the continuum dislocation theory a polycrystal
is thus a stress-free configuration of dislocations. In 
section~\ref{sec:StressFree}, we derive
the most general solution for the dislocation density tensor with zero stress.
In section~\ref{sec:FlatWallDecomposition}, we show that this solution can
be decomposed into a
superposition of flat grain boundaries. In appendix~\ref{app:examples} and 
figure~\ref{fig:circularwall} we explicitly represent a (zero-stress)
curved grain boundary as a continuum superposition of flat walls.
In section~\ref{sec:Rotation}, we also show that the most general stress-free
state can be 
represented in terms of a local rotation field. Hence the stress-free
states in the continuum theory can be interpreted as polycrystals with 
arbitrarily small crystallites; the strain energy for grain boundaries
is zero, so the crystalline axes can vary arbitrarily in space. This would
suggest that sharp, discrete wall formation is not implied by the energetics
within the continuum theory; recent work 
elsewhere~\cite{LimkSeth06,LimkSeth06b} has shown that they may nonetheless
form via shock formation in the natural dislocation dynamics.

At low temperatures, where mass transport by
diffusion of vacancies and interstitial diffusion is frozen out, external
strain is relieved by (volume-preserving) dislocation glide. The motion,
entanglement, and multiplication of dislocations under this low-temperature
plastic deformation leads to work hardening and the development of a yield
stress. These systems have long been modeled using continuum dislocation 
theories,\cite{LimkSeth06,LimkSeth06b,Grom97,%
BakoGrom99,ZaisMiguGrom01,ElAz00,RoyAcha05,AchaSawa06,AchaRoy06,RoyAcha06}
sometimes simplifying the dislocation density into a scalar
quantity, but sometimes incorporating information not contained in the 
Nye tensor (`geometrically unnecessary' dislocation densities
with canceling topological charge, yield stress laws, and separate dislocation
densities for each slip system, \dots). 
Continuum dislocation theories have been important not only in understanding
large-scale discrete dislocation 
simulations,\cite{MiguVespZappWeis01,BakoGrom99,BartCarl97,GullHart93,%
GromPawl93PMA,GromPawl93MSEA,FourSala96,BenzBrecNeed04,BenzBrecNeed05,%
GromBako00,GomeDeviKubi06} but also in understanding emergent
dislocation avalanche phenomena.\cite{ZaisMiguGrom01,%
MiguVespZappWeis01,MiguVespZappWeis01b,MiguVespZaisZapp02,WeisMars03,%
RichWeisLouc05,Samm05,MiguZapp06,Zais06,DimiWoodLeSa06}
At large deformations (stage~III plasticity), these tangles of 
dislocations begin to organize again into walls (here called {\em cell walls})
separating largely undeformed regions\cite{HughChrzLiuHans98,WilsHans91,%
HughHans93,HughHans95,HughHans01,Hugh01} with small misorientation angles
between cells (again potentially explained by a shock formation in the 
climb-free dynamical evolution law~\cite{LimkSeth06,LimkSeth06b}).
In section~\ref{sec:GetLambda} we provide the relation between the local
rotation field for a general dislocation density tensor, which (for general
stressful densities) is a formula providing a least-squares best 
approximation for the local orientation of the crystalline axes.

\section{The general stress-free dislocation density}
\label{sec:StressFree}
A dislocation is a crystallographic defect representing extra rows or
columns of atoms and is characterized by two quantities; its line
direction $\vect{t}$, and its Burgers vector $\vect{b}$. After a
passage around a closed contour $C$ that encircles a dislocation
line, a displacement field $\vect{u}$ receives an increment $\vect{b}$
according to
\begin{equation}\label{E:Burgers}
  \oint_C du_i = \oint_C \betaP_{ij}\,dx_j = -b_i
\end{equation}
where $\tensor{\betaP}$ describes the irreversible plastic deformation.
The Nye dislocation density tensor $\tensor{\rho}$ is defined by
$\tensor{\rho} = (\vect{t}\otimes\vect{b})\delta(\vectsym{\xi})$ where
$\vectsym{\xi}$ is the two-dimensional radius vector measured from the
axis of the dislocation in the plane locally perpendicular to
$\vect{t}$. When many dislocations labeled by $\alpha$ are present, a
coarse-graining description of a conglomerate of dislocations is
preferred. In this picture,
\begin{equation}
  \rho_{ij}(\vect{x}) = \sum_\alpha\int t^\alpha_i b^\alpha_j
  \delta(\vect{x}' - \vectsym{\xi}^\alpha)G(\vect{x}-\vect{x}')
  d^3\vect{x}',
\end{equation}
with Gaussian weighting $G(\vect{x}-\vect{x}') \simeq
(1/\sqrt{2\pi}L)^3\,\text{exp}[-(\vect{x}-\vect{x}')^2/(2L^2)]$ over
some distance scale $L$ large compared to the distance between
dislocations and small compared to the dislocation structures being
modeled. Since dislocation lines are topological and cannot end
inside the crystal, $\tensor{\rho}$ is divergence free:
$\partial_i\rho_{ij} = 0$.

A dislocation strains the crystal, and creates a long-range stress
field. Peach and Koehler derived the relationship for stress
fields due to dislocations in an isotropic material.\cite{PeacKoeh50}
In terms of the Nye dislocation tensor $\tensor{\rho}$, the stress
can be written as the sum of two convolutions:
\begin{multline}\label{E:stress1}
  \sigma_{\alpha\beta}(\vect{x}) = -\frac{\mu}{8\pi}\int_V \\
  \Big[(\varepsilon_{im\alpha}\rho_{\beta m}(\vect{x}') +
  \varepsilon_{im\beta}\rho_{\alpha m}(\vect{x}')) \frac{\partial^3
    |\vect{x}-\vect{x}'|}{\partial x'_i \partial x'_j \partial
  x'_j} 
  -\frac{\mu}{4\pi(1-\nu)} \\ \varepsilon_{imk}
  \rho_{km}(\vect{x}') \left(
    \frac{\partial^3 |\vect{x}-\vect{x}'|}{\partial x'_i \partial
  x'_{\alpha} \partial
      x'_{\beta}} - \delta_{\alpha\beta}\frac{\partial^3
  |\vect{x}-\vect{x}'|}{\partial
      x'_i \partial x'_j \partial x'_j} \right)\Big]\,
  d^3\vect{x}'.
\end{multline}
In Fourier space, the stress is given as a product:
\begin{equation}\label{E:stress_fourier2}
  \Fourier{\sigma}_{\alpha\beta}(\vect{k}) =
  K_{\alpha\beta\mu\nu}(\vect{k})\rhoF_{\mu\nu}(\vect{k})\,,
\end{equation}
where the kernel
\begin{multline*}
  K_{\alpha\beta\mu\nu}(\vect{k}) =  -\frac{i\mu
    k_\gamma}{k^2} \\ \left[
    \varepsilon_{\gamma\nu\alpha}\delta_{\beta\mu} +
    \varepsilon_{\gamma\nu\beta}\delta_{\alpha\mu} +
    \frac{2\varepsilon_{\gamma\nu\mu}}{1-\nu}\left(
      \frac{k_{\alpha}k_{\beta}}{k^2} - \delta_{\alpha\beta} \right)
  \right].
\end{multline*}

The problem of finding a family of dislocation configurations with
zero stress is equivalent to finding those densities
$\Fourier{\tensor{\rho}}$ which are divergence free ($ik_i
\Fourier{\rho}_{ij} = 0$) and are in
the null space of $\tensor{K}$.
Systematic investigation using \emph{Mathematica}$^\circledR$ shows
that the solutions to the system of equations which
incorporate both setting $K_{ijkm}\rhoF_{km}=0$ and $ik_i\rhoF_{ij}
= 0$ are
\begin{alignat}{2}
  &\rhoF_{xx} =
  -\frac{k_y}{k_z}\rhoF_{yz}-\frac{k_z}{k_y}\rhoF_{zy}\,, \quad
  &&\rhoF_{yy} = 
  -\frac{k_x}{k_z}\rhoF_{xz}-\frac{k_z}{k_y}\rhoF_{zy}\,, \notag \\
  &\rhoF_{zz} =
  -\frac{k_x}{k_z}\rhoF_{xz}-\frac{k_y}{k_z}\rhoF_{yz}\,,
  &&\rhoF_{xy} = \frac{k_y}{k_z}\rhoF_{xz}\,, \notag \\
  &\rhoF_{yx} =
  \frac{k_x}{k_z}\rhoF_{yz}\,, \quad &&\rhoF_{zx} =
  \frac{k_x}{k_y}\rhoF_{zy}\,,
\end{alignat}
or, in matrix form,
\begin{equation}
  \tensor{K}'\cdot\vectsym{\rho} = 
  \begin{pmatrix}
    1 & 0 & 0 & 0 & 0 & \dfrac{k_y}{k_z} & 0 & \dfrac{k_z}{k_y} & 0 \\
    0 & 0 & \dfrac{k_x}{k_z} & 0 & 1 & 0 & 0 & \dfrac{k_z}{k_y} & 0 \\
    0 & 0 & \dfrac{k_x}{k_z} & 0 & 0 & \dfrac{k_y}{k_z} & 0 & 0 & 1 \\
    0 & 1 & -\dfrac{k_y}{k_z} & 0 & 0 & 0 & 0 & 0 & 0 \\
    0 & 0 & 0 & 1 & 0 & -\dfrac{k_x}{k_z} & 0 & 0 & 0 \\
    0 & 0 & 0 & 0 & 0 & 0 & 1 & -\dfrac{k_x}{k_y} & 0 
  \end{pmatrix}
  \begin{pmatrix}
    \rhoF_{xx}\\ \rhoF_{xy}\\ \rhoF_{xz}\\ \rhoF_{yx}\\ \rhoF_{yy}\\
    \rhoF_{yz}\\ \rhoF_{zx}\\ \rhoF_{zy}\\ \rhoF_{zz}
  \end{pmatrix} = \vectsym{0}
\end{equation}
Since any given $\tensor{\rhoF}$ has nine 
components and six constraints, and since (\emph{e.g.} for
$\vect{k}=0$) $\tensor{K}'$ has full rank, three basis tensors span
the space of solutions. We solve for them and label them $\tensor{E}^x,
\tensor{E}^y$ and $\tensor{E}^z$.
\begin{equation*}
  \tensor{E}^x = \begin{pmatrix} 0 & ik_y & ik_z \\ 0 & -ik_x & 0 \\
    0 & 0 & -ik_x \end{pmatrix},\,
  \tensor{E}^y = \begin{pmatrix} -ik_y & 0 & 0 \\ ik_x & 0 & ik_z \\
    0 & 0 & -ik_y \end{pmatrix},\,
\end{equation*}
\begin{equation}
  \tensor{E}^z = \begin{pmatrix} -ik_z & 0 & 0 \\ 0 & -ik_z & 0 \\
    ik_x & ik_y & 0 \end{pmatrix},
\end{equation}
or simply
\begin{equation}\label{E:stressfree_basis}
  \boxed{E^{\alpha}_{ij} = -ik_{\alpha}\delta_{ij} +
    ik_j\delta_{i\alpha} 
    = ik_l\varepsilon_{ilm}\varepsilon_{j\alpha m}}
\end{equation}
Direct substitutions of the form of $\tensor{E}^{\alpha}$ in place
of $\tensor{\rhoF}$ show that (\ref{E:stress_fourier2}) and
the divergence free condition are 
simultaneously satisfied for all values of $\alpha$. It is convenient
to include the imaginary number $i$ into the expression for
$\tensor{E}^{\alpha}$, because the Fourier transform of the gradient
of a function is given by multiplying by $i \vect{k}$.

A general stress-free dislocation configuration therefore can be
written as a superposition of the three basis tensors
$\tensor{E}^\alpha$:
\begin{equation}
  \Fourier{\rho}^{\text{SF}}_{ij} = E^{\alpha}_{ij}\Fourier{\Lambda}^\alpha
\end{equation}
The coefficients $\Fourier{\Lambda}^l(\vect{k})$ form a valid vector
field (\emph{i.e.} $\Fourier{\vectsym{\Lambda}}$ transforms like a
vector). This vector will play a special role in determining the grain
orientation inside each cell.

\section{Decompositions of a stress free state into flat Frank walls}
\label{sec:FlatWallDecomposition}
These three basis tensors can be used to describe grain boundaries. As
an example, consider a 
tilt boundary in the $x$-$y$ plane constructed from a set of
parallel dislocation lines pointing along the $\hvect{x}$ direction
with the Burgers vector
$\vect{b}$ pointing along the $\hvect{z}$ direction. Let $n$ be the
number of dislocation lines per unit length along $\hvect{y}$. To make
a plane in real space, we need two $\delta$-functions in
Fourier space. The boundary is then written
\begin{equation}
\begin{split}
  \tensor{\rhoF}^{\text{tilt}\,\hvect{x}} &=
  (2\pi)^2 \frac{nb}{ik_z}\,\delta(k_x)\delta(k_y)\tensor{E}^x \\
  &=
  (2\pi)^2nb\,\delta(k_x)\delta(k_y)\begin{pmatrix} 0&0&1\\0&0&0\\0&0&0
  \end{pmatrix}\!.
\end{split}
\end{equation}
Notice that, for low-angle boundaries (small $n$), the tilt
misorientation angle about the $\hvect{x}$
axis is given by $\omega_x = nb$.

We can write this tilt boundary in
terms of our stress-free basis function $\tensor{E}^x$. But why is the
tilt-boundary stress free? Real grain boundaries have stresses from
their constituent dislocations that cancel at long distances---they
decay exponentially with distance over a length scale given by a
typical distance between dislocations. Hence in the continuum limit
where the dislocations become infinitely close together, the stress
vanishes. Equivalently, the elastic energy of a boundary with low
misorientation angle $\theta$ goes as $-b\,\theta\log\theta$, which
vanishes in the continuum limit $b\rightarrow 0$. Grain boundaries
mediating rigid rotations have vanishing stress in the mesoscale
continuum dislocation theory.

Similarly, a twist boundary in the $x$-$y$ plane can be generated by
two sets of parallel dislocations oriented perpendicular to one
another, one pointing in the $\hvect{x}$ direction while another
pointing in the $\hvect{y}$ direction. It can be written simply as
\begin{equation}\label{E:twistBoundary}
  \begin{split}
    \tensor{\rhoF}^{\text{twist}} &=
    -(2\pi)^2\frac{nb}{ik_z}\,\delta(k_x)\delta(k_y)\tensor{E}^z \\
    &=
    (2\pi)^2 nb\,\delta(k_x)\delta(k_y)\begin{pmatrix}
    1&0&0\\0&1&0\\0&0&0 \end{pmatrix}\!, \\
    &=
    (2\pi)^2 nb\,\delta(k_x)\delta(k_y)\!\!\left[\!\! 
      \begin{pmatrix}1\\0\\0
      \end{pmatrix} \!\otimes \!
      \begin{pmatrix}1\\0\\0\end{pmatrix}\!
      + \!\begin{pmatrix}0\\1\\0\end{pmatrix} \!\otimes \!
      \begin{pmatrix}0\\1\\0\end{pmatrix}\!\!\right]\!,
  \end{split}
\end{equation}
with the twist misorientation angle $\omega_z = nb$.
The fact that one needs two perpendicular sets of parallel
dislocations comes out naturally in this formulation. Because the
number densities of the screw dislocations are the same in both
directions, $n$ here denotes the number density in one of the two
directions.

\begin{figure}[htb]
  \centering
  \psfrag{q}{$\vectsym{\omega}$}
  \psfrag{x}{$\hvect{x}$}
  \psfrag{y}{$\hvect{y}$}
  \psfrag{z}{$\hvect{z}$}
  \psfrag{n}{$\hvect{n}$}
  \psfrag{t}{$\theta$}
  \psfrag{f}{$\phi$}
  \psfrag{D}{$\vectsym{\Delta}$}
  \includegraphics[width=3in]{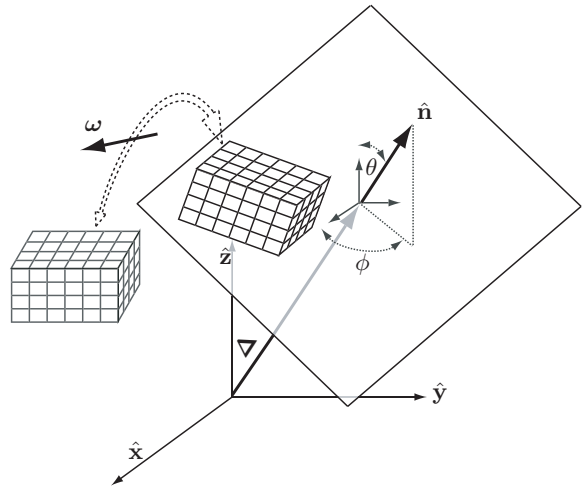}
  \caption{
    A general grain boundary whose normal is $\hvect{n}$
    positioned at the distance $\vectsym{\Delta}$ away from the
    origin separates two unstrained regions
    with a relative orientation defined by $\vectsym{\omega}$.}
  \label{fig:GB_orientation}
\end{figure}%
A general boundary on the $x$-$y$ plane is the sum of three types of
boundary (a tilt along $\hvect{x}$, a tilt along $\hvect{y}$, and a
twist along $\hvect{z}$):
\begin{equation}\label{E:xyBoundary}
  \tensor{\rhoF}^{x\text{-}y} = \tensor{\rhoF}^{\text{tilt}\,\hvect{x}} +
  \tensor{\rhoF}^{\text{tilt}\,\hvect{y}} + \tensor{\rhoF}^{\text{twist}} =
  (2\pi)^2 \frac{\delta(k_x)\delta(k_y)}{ik_z}\,\omega_n\tensor{E}^n\,,
\end{equation}
where $\vectsym{\omega}$ is the Rodrigues vector giving the angle of
misorientation across the wall. The wall can be translated to a new
position ($z = \Delta$) by multiplying by $\exp{-ik_z\Delta}$. The
most general grain boundary with an arbitrary plane orientation can be
obtained by then further rotating equation~\ref{E:xyBoundary} by the
rotation matrix
\begin{multline}\label{E:TheRotation}
  \tensor{R}^{-1}[\Omega=(\theta,\phi)] =
    \left[\tensor{R}_{\hvect{z}}(\phi)\cdot
      \tensor{R}_{\hvect{y}}(-\theta) \right]^{-1} \\
    = 
    \begin{pmatrix}
      \cos(\theta)\cos(\phi) & \cos(\theta)\sin(\phi) & -\sin(\theta) \\
      -\sin(\phi) & \cos(\phi) & 0 \\
      \sin(\theta)\cos(\phi) & \sin(\theta)\sin(\phi) & \cos(\theta)
    \end{pmatrix},
\end{multline}
to get
\begin{equation}\label{E:rhoGBFourier}
  \Fourier{\rho}^{\text{GB}}_{ij}
  [\vect{k},\vectsym{\omega},\Omega,\vectsym{\Delta}] =
  (2\pi)^2\,\frac{\delta(R^{-1}_{xp}k_p)
    \delta(R^{-1}_{yq}k_q)}{iR^{-1}_{zr}k_r}
  \,\omega_n E^n_{ij}\,
  \exp{-i\vect{k}\cdot\vectsym{\Delta}}\,,
\end{equation}
where $\Omega \Leftrightarrow (\theta,\phi)$ defines a unit vector
$\hvect{n}$ normal to the plane of the boundary (see
figure~\ref{fig:GB_orientation}), and $\vect{k}\cdot\vectsym{\Delta} =
k_i n_i\Delta = k_i R_{iz}\Delta =
R^{-1}_{zi}k_i\Delta$. Equation~\ref{E:rhoGBFourier} is Frank's
formula in the language of continuum dislocations. The
connection with Frank's original formula
is discussed in appendix~\ref{app:FrankCondition}.

To take this one step further, since it is possible to decompose
any stress-free state into a linear combination of the tensor
$\tensor{E}^\alpha$, it should also be possible to write a
stress-free state as a superposition of flat cell walls.
\begin{theorem}\label{thm:decomp}
  Any stress-free state $\tensor{\rhoF}^{\text{SF}}$ can be
  written as a 
  superposition of flat cell walls. Or more precisely,
  \begin{multline}\label{E:rhoF_decomp}
    \rhoF^{\text{SF}}_{ij}(\vect{k}) =
    E^\alpha_{ij}\,\Fourier{\Lambda}^\alpha(\vect{k}) = \\
    \int_{-\infty}^{\infty}\!
    d\Delta \int\! d\Omega
    \int\! d^3\vectsym{\omega}\,\left(
      a[\vectsym{\omega},\Omega,\Delta]\cdot
      \rhoF^{\text{GB}}_{ij}[\vect{k},
      \vectsym{\omega},\Omega,\vectsym{\Delta}]\right),
  \end{multline}
  where $\rhoF^{\text{GB}}_{ij}$ is as previously defined, and
  \begin{multline}\label{E:a}
    a[\vectsym{\omega},\Omega,\Delta] =
    \frac{i\,\omega_l}{(2\pi)^3 \pi^{\nicefrac{3}{2}}}\,
    \exp{-|\vectsym{\omega}|^2}\, \int_{-\infty}^{\infty} dk' \,k'^3
    \exp{ik'\Delta} \\
    \Fourier{\Lambda}^l\left[\{k'\sin(\theta)\cos(\phi),
      k'\sin(\theta)\sin(\phi),k'\cos(\theta)\}\right].
  \end{multline}
\end{theorem}
\begin{proof}
To get a general stress-free dislocation distribution, one needs
to integrate over three parameters denoting the misorientation
between the two grains, two angles defining each boundary, and the
position of each grain component.

To show this, we substitute the form of
$a[\vectsym{\omega},\Omega,\Delta]$ into equation~\ref{E:rhoF_decomp}.
\begin{equation}\label{E:pf}
  \begin{split}
    \rhoF^{\text{SF}}_{ij}(\vect{k}) &=
    \int_{-\infty}^{\infty}\! d\Delta \int\! d\Omega  \int\!
    d^3\vectsym{\omega}\, 
    \frac{i}{(2\pi)^3 \pi^{\nicefrac{3}{2}}}\omega_l\,
    \exp{-|\vectsym{\omega}|^2} \\
    &\qquad\times\int_{-\infty}^{\infty} dk'\, k'^3
    \Fourier{\Lambda}^l\left[\ldots\right]\exp{ik'\Delta}
    \\
    &\qquad\qquad \times (2\pi)^2\,\frac{\delta(R^{-1}_{xp}k_p)
      \delta(R^{-1}_{yq}k_q)}{iR^{-1}_{zr}k_r}\,\omega_n E^n_{ij}\,
    \exp{-i\vect{k}\cdot\vectsym{\Delta}} \\
    &= \frac{E^n_{ij}}{4\pi} \int_{-\infty}^{\infty}\!
    dk' {k'}^3 \Fourier{\Lambda}^l\left[\ldots\right] \int\! d\Omega\,
    \frac{\delta(R^{-1}_{xp}k_p)
      \delta(R^{-1}_{yq}k_q)}{R^{-1}_{zr}k_r} \\
    &\underbrace{
      \frac{2}{\pi^{\nicefrac{3}{2}}}
        \int\! d^3\vectsym{\omega}\,
        \omega_l\omega_n\,\exp{-|\vectsym{\omega}|^2}
    }_{\delta_{ln}}    
    \underbrace{\int_{-\infty}^{\infty} d\Delta\,
      \exp{i(k' -
        R^{-1}_{zr}k_r)\Delta}}_{2\pi\delta(k'-R^{-1}_{zr}k_r)} \\
    &= \frac{E^l_{ij}}{2} \int d\Omega \,\delta(R^{-1}_{xp}k_p)
    \delta(R^{-1}_{yq}k_q)\,(R^{-1}_{zr}k_r)^2
    \Fourier{\Lambda}^l[\ldots]
  \end{split}
\end{equation}
The integral over solid angle vanishes except along the line defined
by the product of the two $\delta$-functions. Since our problem is
isotropic, we may take this to be along the $k_x$ direction without
loss of generality. Then the integral reduces to
\begin{equation}
  \begin{split}
    &\int_0^{2\pi}\! d\phi \int_{-1}^1 \! d(\cos\theta) \\
    &\qquad\qquad\delta(k_x\cos\theta)\delta(k_x\sin\phi)
    \underbrace{(k_x\sin\theta\cos\phi)^2}_{
    (k_x\cos\phi)^2(1-\cos^2\theta)}\Fourier{\Lambda}^l[\dots] \\
    &\qquad = \int_0^{2\pi}\!
    d\phi\,\delta(k_x\sin\phi)|k_x\cos\phi|
    \Fourier{\Lambda}^l[\{k_x\cos^2\phi,0,0\}] \\
    &\qquad = \int_0^{2\pi}\! d\phi \left(
    \frac{\delta(\phi-0)}{|k_x\cos\phi|} +
    \frac{\delta(\phi-\pi)}{|k_x\cos\phi|}\right) \\
    &\qquad\qquad\qquad\qquad\qquad\quad|k_x\cos\phi|
    \Fourier{\Lambda}^l[\{k_x\cos^2\phi,0,0\}] \\
    &\qquad = 2\Fourier{\Lambda}^l[\{k_x,0,0\}],
  \end{split}
\end{equation}
where we use $\delta(g(x)) = \sum_{a}\delta(x-a)/|g'(a)|$,
 and the sum is taken over all $a$'s with $g(a)=0$ and $g'(a)\ne
 0$.\cite{ArfkWebe95} The argument works for any $\vect{k}$. Thus
    $\rhoF^{\text{SF}}_{ij}(\vect{k}) =
    E^\alpha_{ij}\,\Fourier{\Lambda}^\alpha(\vect{k})$ is shown.
\end{proof}

One must emphasize that this theorem does not explain the prevalence
of grain boundaries. Most stress-free states will be formed by
continuous superpositions of walls. Indeed, even a curved grain
boundary will demand such a continuous superposition (see
appendix~\ref{app:examples}).

\section{Stress-free states and continuous rotational deformations}
\label{sec:Rotation}
In this section we show that the vector field
$\vectsym{\Lambda}(\vect{x})$ introduced in the previous section
is precisely the Rodrigues vector field giving the rotation matrix
that describes the local orientation of the crystalline axes at
position $\vect{x}$.

What is $\vectsym{\Lambda}^{\text{GB}}(\vect{x})$ associated with a
grain boundary? Consider the form of
$\vectsym{\Fourier{\Lambda}}^{xy}$ for a 
boundary lying in the $x$-$y$ plane:
\begin{equation}
  \vectsym{\Fourier{\Lambda}}^{x\text{-}y}(\vect{k}) = (2\pi)^2
  \frac{\delta(k_x)\delta(k_y)}{ik_z}\,\vectsym{\omega}
\end{equation}
The inverse Fourier transform of this expression involves an integral
over a semi-circular contour in the upper complex plane, resulting in
\begin{equation}
  \vectsym{\Lambda}^{x\text{-}y}(\vect{x}) =
  \frac{1}{2}\,\text{sign}[z]\vectsym{\omega}.
\end{equation}
In general, $\vectsym{\Lambda}^{\text{GB}}$ is found after proper
translation and rotation of the plane:
\begin{equation}
  \vectsym{\Lambda}^{\text{GB}}(\vect{x}) =
  \frac{1}{2}\,\text{sign}[\hvect{n}\cdot(\vect{x}-
  \vectsym{\Delta})]\vectsym{\omega}\,.
\end{equation}
The vector $\vectsym{\Lambda}(\vect{x})$ provides
information about the local crystal orientation at the point
$\vect{x}$ relative to a fixed global orientation. This is true in
general:
\begin{theorem}
  The direction of
  $\vectsym{\Lambda}$ gives the
  axis of rotation of the local crystal orientation with respect to
  a fixed global coordinates by the amount provided by its
  magnitude.
\end{theorem}
In other words, the Rodrigues vector
$\vectsym{\Lambda}(\vect{x})$ describes the local crystal
orientations due to the presence of the stress-free dislocation
density field $\tensor{\rho}^{\text{SF}}$.
\begin{proof}
First, note that
\begin{equation}
  \Fourier{\rho}^{\text{SF}}_{ij} =
  E^\alpha_{ij}\,\Fourier{\Lambda}^\alpha = ik_j\Fourier{\Lambda}_i -
  \delta_{ij} ik_m\Fourier{\Lambda}_m\,,
\end{equation}
which, in real space, corresponds to
\begin{equation}\label{E:rhoLambda}
  \rho^{\text{SF}}_{ij} = \partial_j\Lambda_i -
  \delta_{ij}\partial_m\Lambda_m.
\end{equation}

Now consider a rotation field $R_{ij}(\vectsym{\Lambda}') =
\exp{\varepsilon_{jik}\Lambda'_{k}}$ where
$\tensor{\Lambda}'(\vect{x})$ is the 
Rodrigues vector giving the local orientation, and we wish to argue
that $\tensor{\Lambda}'$ can be used for $\tensor{\Lambda}$. Consider
a small Burgers circuit $C$ enclosing a region $S$ with local
orientation given by the field of
$\vectsym{\Lambda}'(\vect{x})$.
Integrating around the circuit $C$, the net closure failure
$-\vect{b}$ due to the plastic distortion $\tensor{\betaP}$ is given in
terms of the local rotation $\tensor{R}(\vectsym{\Lambda}')$ (see
equation~\ref{E:Burgers}):
\begin{equation}\label{E:closureFailure}
  -b_j = \oint_C \betaP_{ij}\,dx_i = \oint_C
   R_{ji}(\vectsym{\Lambda}'(\vect{x}))\,dx_i
\end{equation}
Applying Stokes' theorem to
equation~\ref{E:closureFailure} and noting that the change in
$\vectsym{\Lambda}'$ is small inside the small circuit $C$, we obtain
\begin{equation}
  \begin{split}
    \oint_C R_{ji}(\vectsym{\Lambda}'(\vect{x}))dx_i &= \int_S
    \varepsilon_{iln}\partial_l(\exp{\varepsilon_{jnm}\Lambda'_m})dS_i
    \\
    &\simeq \int_S \varepsilon_{iln}\partial_l(\delta_{jn} +
    \varepsilon_{jnm} \Lambda'_m) dS_i \\
    &= \int_S \varepsilon_{iln}\varepsilon_{jnm} \Lambda'_m dS_i =
    -\int_S \rho_{ij}dS_i\,,
  \end{split}
\end{equation}
where we use the definition of the Nye tensor in the last
equality. This expression holds regardless of the enclosed surface
$S$, thus%
\footnote{
  Nye provided the
  relationships between the dislocation density tensor
  $\tensor{\rho}$ and the lattice curvature tensor
  $\tensor{\kappa}$.\cite{Nye53} Let $d\phi_i$ be small lattice rotations
  about three coordinate axes, associated with the displacement
  vector $dx_j$, then $\kappa_{ij} \equiv \partial\phi_i/\partial x_j$.
  He shows that given a curvature tensor $\tensor{\kappa}$ the
  Nye dislocation tensor $\tensor{\rho}$ can be determined,
  \emph{i.e.}, $\rho_{ij} = \kappa_{ij} - \delta_{ij}\kappa_{kk}$.
  By comparing equation~\ref{E:rhoLambda} with the above expression
  we can identify $\partial_j\Lambda_i$
  with the lattice curvature tensor $\kappa_{ij}$ in the stress-free
  regions.
}
\begin{equation}
  \rho_{ij} = -\varepsilon_{iln}\varepsilon_{jnm} \Lambda'_m
  = \partial_j\Lambda'_i - \delta_{ij}\partial_m\Lambda'_m. \qedhere
\end{equation}
\end{proof}
Thus the stress-free distortions are precisely those generated by
rotation fields, and its dislocation density tensor field is given by
our decomposition (equation~\ref{E:rhoLambda}) with
$\vectsym{\Lambda}$ equal to the Rodrigues vector for the local
rotation.

\section{Extracting the local misorientation from the Nye
  tensor}\label{sec:GetLambda}
The decomposition of $\rhoF^{\text{SF}}_{ij} =
\Fourier{\Lambda}^\alpha E^\alpha_{ij}$ is somewhat different from the
problem of breaking up a vector into projections on various basis
vectors. The main distinction lies in the fact that the three
$E^{\alpha}_{ij}$'s are not orthogonal to one another, so finding the
components along them is not a simple dot product. We instead will
minimize the square of the difference between the
actual $\rhoF^{\text{SF}}_{ij}$ and the decomposition
$E^\alpha_{ij}\,\Fourier{\Lambda}^\alpha$. Let's define
\begin{equation}
  f \equiv \sum_{ij}\left(\rhoF_{ij} -
    E^\alpha_{ij}\,\Fourier{\Lambda}^\alpha \right)^2.
\end{equation}
Minimizing $f$ will not only give the correct
$\Fourier{\tensor{\Lambda}}^\alpha$ for a stress-free
$\Fourier{\tensor{\rho}}^{\text{SF}}$, it will also provide a natural
definition for the local crystalline orientation of a general
(stressful) dislocation density field.

The minimization occurs when the derivative with respect to
the component $\Fourier{\Lambda}^\beta$ is zero:
\begin{align}
  \begin{split}
    0 &= \pdif{f}{\Fourier{\Lambda}^\beta} \\
    &= \pdif{\phantom{f}}{\Fourier{\Lambda}^\beta}
    \sum_{ij}\left(\rhoF_{ij} -
    E^\alpha_{ij}\,\Fourier{\Lambda}^\alpha \right)^2 \\
    &= -2 E^\beta_{ij}\left(\rhoF_{ij} -
      E^\alpha_{ij}\,\Fourier{\Lambda}^\alpha \right) \\
    E^\beta_{ij}\,\rhoF_{ij} &= \underbrace{E^\beta_{ij}
      \,E^\alpha_{ij}}_{M_{\alpha\beta}}\, \Fourier{\Lambda}^\alpha,
  \end{split}
\end{align}
or, 
\begin{equation}
  \Fourier{\Lambda}^\alpha =
  M^{-1}_{\alpha\beta}\,E^\beta_{ij}\,\rhoF_{ij}
\end{equation}
where
\begin{equation*}
\begin{split}
  M^{-1}_{\alpha\beta} &= \frac{1}{2k^4}(k_\alpha k_\beta -
  2k^2 \delta_{\alpha\beta}) \\ &= \frac{1}{2k^4}
  \begin{pmatrix}
    k^2_x-2k^2 & k_x k_y & k_x k_z \\
    k_x k_y & k^2_y-2k^2 & k_y k_z \\
    k_x k_z & k_y k_z & k^2_z-2k^2
  \end{pmatrix}
\end{split}
\end{equation*}
and $k^2 \equiv |\vect{k}|^2$.

It is possible to directly compute $\vectsym{\Lambda}$ in real
space. From
\begin{equation}\label{E:Lambda_rhoSF}
  \begin{split}
    \Fourier{\Lambda}^i &=
    M^{-1}_{ij}E^j_{mn}\Fourier{\rho}_{mn} \\
    &= \frac{1}{2k^4}\left[k_i k_j - 2k^2\delta_{ij}
    \right]\left[-ik_j\delta_{mn} + ik_n\delta_{jm}
    \right]\Fourier{\rho}_{mn} \\
    &= \frac{1}{2k^4}\left[ik_i k_m k_n + 2ik^2k_i\delta_{mn} - 2ik^2
      k_n\delta_{im} \right]\Fourier{\rho}_{mn} \\
    &= \frac{i}{k^2}\left[ \frac{k_ik_mk_n}{2k^2} +
      k_i\delta_{mn} - k_n\delta_{im}
    \right]\Fourier{\rho}_{mn} \\
    &= \frac{i}{k^2}\left[k_i\Fourier{\rho}_{nn} -
      k_n\Fourier{\rho}_{in}
    \right].
  \end{split}
\end{equation}
The expression of the Rodrigues vector $\vectsym{\Lambda}$ in real
space, therefore by analogy to $1/k^2$ factor in the Coulomb
potential, is therefore
\begin{equation}\label{E:Lambda_rho}
  \Lambda^{i}(\vect{x}) = \frac{1}{4\pi}\int\frac{
    \partial'_n\rho_{in}(\vect{x}') -
    \partial'_i\rho_{nn}(\vect{x}')}{|\vect{x} -
    \vect{x}'|}\,d^3\vect{x}'.
\end{equation}
Equation~\ref{E:Lambda_rho} should be viewed as a natural definition
of the local crystal axes, which could be invaluable for extracting
information about the misorientation
angle distribution, the wall positions, and hence the grain and
cell size distributions.\cite{SethCoffDeml03}

\section{Conclusions}
In this manuscript, we explored the space of stress--free dislocation
densities for an isotropic system. We showed, from first principles, that
any stress--free state can be decomposed into a superposition of flat
walls (grain boundaries) and also can be written as a local
rotational deformation field $\vectsym{\Lambda}(\vect{x})$. Finally, we
provide a relationship between this rotation field and the Nye
dislocation density tensor, which in addition provides a formula for the best
least-squares approximation for the rotation field for a stressful dislocation
density.

The analysis presented here forms the mathematical framework on which 
dynamical theories of continuum dislocation evolution are hung. It
should offer basic tools for interpreting
these simulations (identifying walls, mis-orientations, and 
rotational deformations during the evolution under polycrystalline coarsening
or plastic deformation), for theories based on the Nye tensor or
more microscopic formulations. It should provide also a theoretical basis
for interpreting wall formation in continuum theories; minimizing stress
provides a rationale for continua of walls, but not for discrete, individual
grain boundaries or cell walls.

\appendix
\section{Frank's formula for a general grain boundary: connection to
  continuum theory}
\label{app:FrankCondition}
Frank gave conditions on dislocation density for wall separating two
perfect crystals mis-aligned by a rigid-body rotation.\cite{Fran50}
Our analysis in section~\ref{sec:FlatWallDecomposition} explicitly
generated such walls, leading to a condition
(equation~\ref{E:rhoGBFourier}) on the Nye dislocation 
density tensor. Here we relate Frank's original formulation with
ours. For simplicity, we shall restrict ourselves to a small angle of
misfit $\theta$. For the treatment of large-angle boundaries,
see Ref.~\onlinecite{ReadShoc50}

Let $\vect{V}$ be an arbitrary vector lying in the plane of a grain
boundary, $\vectsym{\omega}$ be an axis defining the relative
rotation between the two grains separated by the boundary whose
magnitude gives the net rotation angle $\theta$, and $\vect{b}$ be
the sum of the Burgers vectors of the dislocations cut by
$\vect{b}$, Frank's formula reads
\begin{equation}\label{E:FrankCondition}
  \vect{b} = \vect{V}\times\vectsym{\omega}\,.
\end{equation}
(See Ref.~\onlinecite{Read53} for the derivation,%
\footnote{There is a sign
  difference between the formula quoted here and that presented
  in Ref.~\onlinecite{Read53}. This is due to the discrepancy in defining
  the Burgers vector.}
and Ref.~\onlinecite{Fran50} for
the formula with an arbitrarily large angle $\theta$.)

\begin{figure}[htb]
  \centering
  \psfrag{o}{$\vectsym{\omega}$}
  \psfrag{V}{$\vect{V}$}
  \psfrag{S}{$S$}
  \psfrag{b}{$\vect{b}$}
  \psfrag{D}{$\vectsym{\Delta}$}
  \psfrag{n}{$\hvect{n}$}
  \psfrag{xa}{$\vect{x}_a$}
  \psfrag{xb}{$\vect{x}_b$}
  \includegraphics[height=3in]{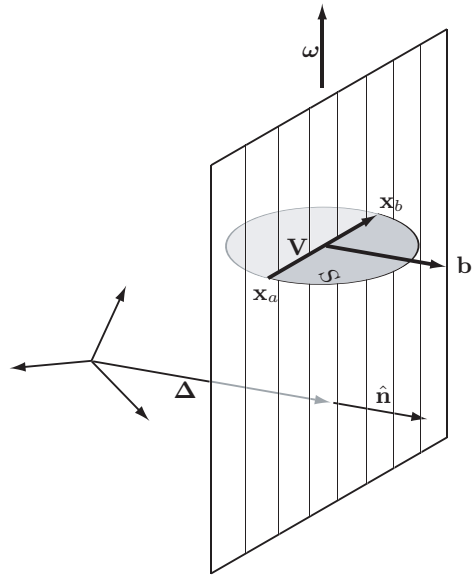}
  \caption{The orientation of the plain is defined by the vector normal
    $\hvect{n}$. The Rodrigues vector $\vectsym{\omega}$ gives the
    axis of rotation and the angle of relative orientation between the
    two grains across the boundary.}
  \label{fig:Frank_wall}
\end{figure}
Using the Nye tensor, we can rephrase (\ref{E:FrankCondition}) and
then compare it with our statement of stress-free
boundaries. Let's start off by defining a Burgers circuit $C$
enclosing a surface $S$ that
intersects a grain boundary at two points $\vect{x}_a$ and
$\vect{x}_b$. The net Burgers vector encompassed by the surface is
$\vect{b}$. Define $\vect{V}$ to be a vector lying in the boundary
plane pointing from $\vect{x}_a$ to $\vect{x}_b$, $\vect{V} \equiv
\vect{x}_b - \vect{x}_a$. We can represent this
grain boundary by a constant matrix $\tensor{\rho}^0$ multiplied
by a plane defined by
$\delta(\hvect{n}\cdot(\vect{x}-\vectsym{\Delta}))$, where
$\hvect{n}$ is a unit vector normal to the plane, and
$\vectsym{\Delta}$ is the perpendicular vector pointing from the
origin to the plane. (See figure~\ref{fig:Frank_wall}.) The
integral of the Nye tensor $\tensor{\rho}$ on the surface $S$
gives the net Burgers vector $\vect{b}$ passing through that surface:
\begin{equation}
  \begin{split}
    b_j &= \int_S \rho_{ij}\,dS_i \\
    &= \int_S
    \rho^0_{ij}\,\delta(\hvect{n}\cdot(\vect{x}
    -\vectsym{\Delta}))[\hvect{n}\times \hvect{V}]_i\,dA
  \end{split}
\end{equation}
The $\delta$-function serves to collapse the area integral
into a line integral since the value is zero outside of the plane
defined by $\vect{x}\cdot\hvect{n} = 0$:
\begin{equation}
  \begin{split}
    b_j &= \int_{\vect{x_a}}^{\vect{x_b}}
    \rho^0_{ij}[\hvect{n}\times\hvect{V}]_i\,dl \\
    &= \rho^0_{ij}\big| \vect{x_b} -
    \vect{x_a}\big|\varepsilon_{imn}\hat{n}_m\hat{V}_n  \\
    &= \rho^0_{ij}\,\varepsilon_{imn}\hat{n}_m V_n
  \end{split}
\end{equation}
We can therefore relate the dislocation density to the rotation
vector $\vectsym{\omega}$ using Frank's formula
(equation~\ref{E:FrankCondition}):
\begin{equation}
  \begin{split}
    \varepsilon_{jpq} \,V_p\,\omega_q &=
    \rho^0_{ij}\,\varepsilon_{imn}\hat{n}_m V_n \\
    0 &= \rho^0_{ij}\,\varepsilon_{imn}\hat{n}_m V_n -
    \varepsilon_{jmn}V_m\omega_n
  \end{split}
\end{equation}
With some relabeling, this becomes
\begin{equation}\label{E:FrankCondition2}
  0 = \left( \rho^0_{ij}\hat{n}_m +
    \delta_{ij}\omega_m\right)\varepsilon_{imn}V_n \,.
\end{equation}

Since $\vect{V}$ is an arbitrary vector
in the plane of the grain boundary, we can write $\vect{V}$ as
$\vect{V} = \hvect{n}\times\vect{W}$ for an arbitrary vector
$\vect{W}$. We can substitute $\hvect{n}\times\vect{W}$ back into
(\ref{E:FrankCondition2}),
\begin{equation}
  0 = \left( \rho^0_{ij}\hat{n}_m +
    \delta_{ij}\omega_m\right)\varepsilon_{imn}
  \varepsilon_{npq}\hat{n}_p W_q\,.
\end{equation}
This condition holds regardless of $\vect{W}$. We can therefore
safely ignore $\vect{W}$ in the equation. The condition now
becomes
\begin{equation}
  \begin{split}
    0 &= \left( \rho^0_{ij}\hat{n}_m +
      \delta_{ij}\omega_m\right)\varepsilon_{imn}
    \varepsilon_{npq}\hat{n}_p \\
    &=\left( \rho^0_{ij}\hat{n}_m +
      \delta_{ij}\omega_m\right)(\delta_{ip}\delta_{mq}
    -\delta_{iq}\delta_{mp}) \hat{n}_p \\
    &= \hat{n}_i\rho^0_{ij}\hat{n}_q + \hat{n}_j\omega_q -
    \rho^0_{qj} - \delta_{qj}\omega_p\hat{n}_p\,.
  \end{split}
\end{equation}
The first term goes to zero because the first index of
$\rho^0_{ij}$ designates the line component which always lies in
the plane of the boundary. By definition, $\hvect{n}$ is
perpendicular to the plane, therefore, $\hat{n}_i\rho^0_{ij} =
0$. The condition for $\tensor{\rho}^0$ that makes a valid grain
boundary is thus
\begin{equation}
  \rho^0_{ij} = \omega_i\hat{n}_j -
  (\vectsym{\omega}\cdot\hvect{n})\delta_{ij}\,,
\end{equation}
or:
\begin{equation}
  \boxed{\tensor{\rho}^{\text{GB}} =
    \left[\vectsym{\omega}\otimes \hvect{n} -
      \left(\vectsym{\omega}\cdot\hvect{n} \right)\tensor{1}
    \right]\delta(\hvect{n}\cdot(\vect{x}-\vectsym{\Delta}))}
\end{equation}

To see the connection between our formalism in obtaining a general
stress-free state, let us again
rewrite the Fourier Transform of the general grain boundary
$\Fourier{\tensor{\rho}}^{\text{GB}}$,
\begin{equation}\tag{\ref{E:rhoGBFourier}'}
  \Fourier{\rho}^{\text{GB}}_{ij} =
  (2\pi)^2\,\frac{\delta(R^{-1}_{xp}k_p)
    \delta(R^{-1}_{yq}k_q)}{iR^{-1}_{zr}k_r}
  \,\omega_n E^n_{ij}\,
  \exp{-i\vect{k}\cdot\vectsym{\Delta}}\,,
\end{equation}
where all the variables are as defined previously.
It is possible to perform the inverse transform of
$\Fourier{\tensor{\rho}}^{\text{GB}}$ to arrive at its real space
representation. The two $\delta$-functions serve to define a plane
in real space. The natural choice of coordinate is to make a
rotational change of variables from $(k_x,k_y,k_z)$ to
$(\xi_x,\xi_y,\xi_z)$ where $\xi_i = R^{-1}_{ij}k_j$. In this
coordinate, $\hat{\vectsym{\xi}}_z$ is perpendicular to the plane
of the boundary. The other two basis vectors lie in the plane of
the boundary.

The inverse transform can be written as
\begin{equation}\label{E:rhoGBrealderiv}
  \begin{split}
    \rho^{\text{GB}}_{ij} &= \frac{1}{(2\pi)^3}\int
    \Fourier{\rho}^{\text{GB}}_{ij}
    \exp{i \vect{k}\cdot\vect{x}} \, d^3\vect{k} \\
    &= \frac{1}{2\pi}\int \frac{\delta(R^{-1}_{xp}k_p)
      \delta(R^{-1}_{yq}k_q)}{iR^{-1}_{zr}k_r}
    \,\omega_n E^n_{ij}\,
    \exp{i\vect{k}\cdot(\vect{x}-\vectsym{\Delta})} \, d^3\vect{k}
    \\
    &= \frac{1}{2\pi}\int \frac{\delta(\xi_x)
      \delta(\xi_y)}{i\xi_z}
    \,\omega_n E^n_{ij}\,
    \exp{i\vect{k}\cdot(\vect{x}-\vectsym{\Delta})} \,
    d^3\vectsym{\xi}\,.
  \end{split}
\end{equation}
Note that since the new basis vectors are the rotation of the
original set, its Jacobian is one. The next step is to express
$E^n_{ij}$ in terms of the new basis:
\begin{equation}
  \begin{split}
    E^n_{ij} &= ik_j\delta_{in} - ik_n\delta_{ij} \\
    &= iR_{jm}R^{-1}_{mp}k_p\delta_{in} - i
    R_{nm}R^{-1}_{mp}k_p\delta_{ij} \\
    &= iR_{jm}\xi_m\delta_{in} - iR_{nm}\xi_m\delta_{ij} \\
    &= i\xi_m\left( R_{jm}\delta_{in} - R_{nm}\delta_{ij}\right)
  \end{split}
\end{equation}
Similarly,
\begin{equation}
  \exp{i\vect{k}\cdot(\vect{x}-\vectsym{\Delta})} =
  \exp{i R_{ij}R^{-1}_{jm}k_m ( x_i - \Delta_i))} = \exp{i
    R_{ij}\xi_j (x_i - \Delta_i)}\,.
\end{equation}
Substituting these into (\ref{E:rhoGBrealderiv}) gives
\begin{equation}
  \begin{split}
    \rho^{\text{GB}}_{ij} &= \frac{1}{2\pi}\int \frac{\delta(\xi_x)
      \delta(\xi_y)}{i\xi_z}
    \,\omega_n i\xi_m \\
    &\qquad\qquad\qquad\left( R_{jm}\delta_{in} -
      R_{nm}\delta_{ij}\right)\, 
    \exp{iR_{ij}\xi_j (x_i - \Delta_i)} \,
    d^3\vectsym{\xi} \\
    &= \frac{1}{2\pi}\int_{-\infty}^{\infty} \!
    \omega_n \left( R_{jm}\delta_{in} -
      R_{nm}\delta_{ij}\right)\, 
    \exp{iR_{iz}\xi_z (x_i - \Delta_i)} \, d\xi_z \\
    &= \left(\omega_i R_{jz} - \omega_n R_{nz}
      \delta_{ij}\right)\underbrace{\frac{1}{2\pi}
      \int_{-\infty}^{\infty}\exp{iR_{iz}\xi_z (x_i - \Delta_i)}
      \, d\xi_z}_{\delta(R_{iz}(x_i-\Delta_i))} \,.
  \end{split}
\end{equation}
The rotation matrix $\tensor{R}$ was so constructed that
$\tensor{R}\cdot\hvect{z} = \hvect{n}$, or $R_{iz} =
\hat{n}_i$. Therefore,
\begin{equation}
  \rho^{\text{GB}}_{ij} = \left[\omega_i\hat{n}_j -
    (\vectsym{\omega}\cdot\hvect{n})\delta_{ij}\right]
  \delta(\hvect{n}\cdot(\vect{x}-\vectsym{\Delta}))\,,
\end{equation}
exactly the same as what we derived from Frank's formula.

\section{Decomposing stress-free states into flat walls: two examples}
\label{app:examples}
Here we illustrate theorem~\ref{thm:decomp} and
equation~\ref{E:rhoGBFourier} by decomposing two stress-free states
into a sum $a[\vectsym{\omega},\Omega,\Delta]$ of flat Frank walls.
Let's start with one of the simplest examples which is a flat twist
boundary. According to equation~\ref{E:twistBoundary} the boundary, in
Fourier space, can be written as
\begin{equation}\tag{\ref{E:twistBoundary}'}
  \tensor{\rhoF}^{\text{twist}} =
  -\frac{nb}{ik_z}\,\delta(k_x)\delta(k_y)\tensor{E}^z.
\end{equation}
The form of 
$a[\vectsym{\omega},\Omega,\Delta]$, according to
equation~\ref{E:a}, in this case is
\begin{equation}
  \begin{split}
    a^{\text{twist}} &=
    \frac{i\omega_z}{(2\pi)^3\pi^{\nicefrac{3}{2}}}\,
    \exp{-|\vectsym{\omega}|^2} \int_{-\infty}^\infty\!
    dk' \\
    &\qquad k'^3\frac{(-nb)\,\delta(k'\sin\theta\cos\phi)
      \delta(k'\sin\theta\sin\phi)}{ik'\cos\theta}\,\exp{ik'\Delta}
    \\
    &= -\frac{nb\,\omega_z
      \exp{-|\vectsym{\omega}|^2}}{(2\pi)^3\pi^{\nicefrac{3}{2}}}
    \frac{\delta(\sin\theta\cos\phi) \delta(\sin\theta
      \sin\phi)}{\cos\theta}\,2\pi\delta(\Delta).
  \end{split}
\end{equation}
The combination of $\delta$-function implies that $\phi=0$ or
$\phi$, thus
\begin{equation}
  a^{\text{twist}} = -\frac{nb\,\omega_z
    \exp{-|\vectsym{\omega}|^2}}{
    (2\pi)^2\pi^{\nicefrac{3}{2}}}\,
  \delta(\cos\phi)\delta(\sin\phi)\delta(\Delta),
\end{equation}
implying that such a wall can be created by only one regular
straight wall.

\begin{figure}[hbt]
  \centering
  \includegraphics[width=3in]{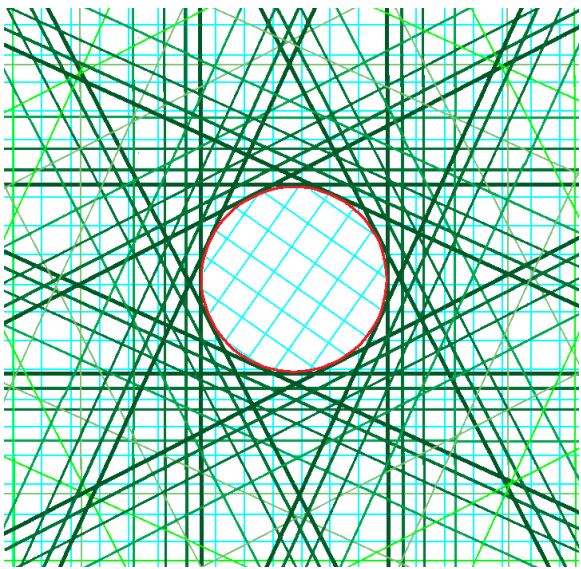}
  \caption{(Color online) {\bf A circular grain boundary} can be
    decomposed into a series of flat walls whose density decays as
    $1/\Delta^3$ away from the center of the cylindrical cell.}
  \label{fig:circularwall}
\end{figure}
A more complicated example is the case where one cuts out a
cylindrical portion of radius $R$ inside a crystal with the axis
of symmetry pointing along $\hvect{z}$, rotates it, and pastes it 
back (figure~\ref{fig:circularwall}). The resulting boundary is a
circular grain boundary which
can be represented in Fourier space as
\begin{equation}
  \tensor{\rhoF}^{\text{circ}} =
  \BesselJ_1\left[\sqrt{k_x^2+k_y^2}\,R \right]
  \frac{\delta(k_z)}{\sqrt{k_x^2 + k_y^2}} \tensor{E}^z,
\end{equation}
where $\BesselJ_1[\cdot]$ is the Bessel function of type 1.
In this case,
\begin{equation}
  \begin{split}
    a^{\text{circ}} &=  \frac{i\omega_z}{(2\pi)^3\pi^{\nicefrac{3}{2}}}\,
    \exp{-|\vectsym{\omega}|^2} \\
    &\qquad\quad\times\int_{-\infty}^\infty\!
    dk'\,k'^3
    \frac{\BesselJ_1\left(|k'\sin\theta|\,R\right)}{|k'\sin\theta|}
    \delta(k'\cos\theta)\,\exp{ik'\Delta}\\
    &= \frac{i\omega_z
    \exp{-|\vectsym{\omega}|^2}}{(2\pi)^3\pi^{\nicefrac{3}{2}}}\,    
    \delta(\cos\theta)
    \underbrace{\int_{-\infty}^{\infty}\!dk'\,|k'|\BesselJ_1(|k'|R)\,
      \exp{ik'\Delta}}_{\frac{-2iR}{(\Delta^2 -
        R^2)^{\nicefrac{3}{2}}}\Theta(\Delta-R)} \\
    &= \frac{2R\,\omega_z\exp{-|\vectsym{\omega}|^2}}%
    {(2\pi)^3\pi^{\nicefrac{3}{2}}(\Delta^2-R^2)^{\nicefrac{3}{2}}}\,
    \delta(\cos\theta) \Theta(\Delta-R).
  \end{split}
\end{equation}
This example emphasizes the important point that we mentioned
earlier, that a stress-free dislocation configuration may need to be
decomposed into a continuous superposition of flat cell walls. In
particular, here we represent a cylindrical wall as an infinite
sum of flat walls with whose amplitudes go down as $1/\Delta^3$
with distance $\Delta$ away from the center of the cylinder.

\begin{acknowledgments}
  We would like to thank Wolfgang Pantleon for suggesting the
  interpretation of $\vectsym{\Lambda}$ in terms of a local rotation
  field, and we acknowledge funding from NSF grants ITR/ASP ACI0085969
  and DMR-0218475.
\end{acknowledgments}

\bibliography{references}

\end{document}